\documentclass[twocolumn]{aastex61}

\begin{document}

\title{T CrB: Radio Observations During the 2016--2017 ``Super-Active'' State}

\author[0000-0002-3873-5497]{Justin D. Linford}
\affiliation{Department of Physics and Astronomy, West Virginia University, P.O. Box 6315, Morgantown, WV 26506, USA}
\affiliation{Center for Gravitational Waves and Cosmology, West Virginia University, Chestnut Ridge Research Building, Morgantown, WV 26505, USA}
\affiliation{National Radio Astronomy Observatory, P.O. Box O, Socorro, NM 87801, USA}

\author[0000-0002-8400-3705]{Laura Chomiuk}
\affiliation{Center for Data Intensive and Time Domain Astronomy, Department of Physics and Astronomy, Michigan State University, East Lansing, MI 48824, USA}

\author{Jennifer L. Sokoloski}
\affiliation{Columbia Astrophysics Laboratory, Columbia University, New York, NY, USA}

\author[0000-0001-7548-5266]{Jennifer H.~S. Weston}
\altaffiliation{AAAS Science and Technology Policy Fellow}
\affiliation{AAAS Fellowship Programs, Inc. (AFPI), American Association for the Advancement of Science, 1200 New York Ave, Washington, DC 20005, USA}

\author{Alexander J. van der Horst}
\affiliation{Department of Physics, The George Washington University, Washington, DC 20052, USA}
\affiliation{Astronomy, Physics, and Statistics Institute of Sciences, The George Washington University, Washington, DC 20052, USA}

\author[0000-0002-8286-8094]{Koji Mukai}
\affiliation{Center for Space Science and Technology, University of Maryland, Baltimore County, Baltimore, MD 21250, USA}
\affiliation{CRESST and X-ray Astrophysics Laboratory, NASA/GSFC, Greenbelt, MD 20771, USA}

\author[0000-0002-8456-1424]{Paul Barrett}
\affiliation{Department of Physics, The George Washington University, Washington, DC 20052, USA}

\author{Amy J. Mioduszewski}
\affiliation{National Radio Astronomy Observatory, P.O. Box O, Socorro, NM 87801, USA}

\author[0000-0003-4197-0524]{Michael Rupen}
\affiliation{Herzberg Institute of Astrophysics, National Research Council of Canada, Penticton, BC, Canada}

\correspondingauthor{Justin D. Linford}
\email{jlinford@nrao.edu}

\received{2019-06-10}
\accepted{2019-08-16}
\submitjournal{The Astrophysical Journal}

\begin{abstract}
We obtained radio observations of the symbiotic binary and known recurrent nova T Coronae Borealis following a period of increased activity in the optical and X-ray bands.  A comparison of our observations with those made prior to 2015 indicates that the system is in a state of higher emission in the radio as well.  The spectral energy distributions are consistent with optically thick thermal bremsstrahlung emission from a photoionized source.  Our observations indicate that the system was in a state of increased ionization in the companion wind, possibly driven by an increase in accretion rate, with the radio photosphere located well outside the binary system.  
\end{abstract}


\section{Introduction}
\label{intro}

T Coronae Borealis (T CrB) is a binary system that wears many different hats.  It is a well-known binary system \citep[e.g.,][]{kg86, fekel2000}.  The system contains a white dwarf and a late-type giant star, with the white dwarf accreting material from the giant.  This qualifies the system as a symbiotic binary \citep[e.g.,][]{allen84}.  The system is also a known recurrent nova, with eruptions in 1866 and 1946 \citep{warner95, schaefer10}.

While there is evidence that the donor star fills its Roche lobe \citep{ym93, bm98}, there is some uncertainty as to the exact spectral type of the donor star.  \cite{bailey75}, \cite{ym93}, and \cite{munari16} identify the donor as a M3III star, while \cite{ms99} and \cite{tcrb_flickering16} identify it as a M4III.  The masses are reported as $1.12 \pm 0.23 M_{\sun}$ for the donor and $1.37 \pm 0.13 M_{\sun}$ for the white dwarf, and the binary orbital inclination is reported as $i \sim 67\degr$ \citep{stan04}.  The binary period is 227.6 days \citep{kg86, fekel2000}, and the orbit is consistent with being circular \citep{fekel2000}.  The binary separation reported by \cite{fekel2000} is $a \sin i = 74.77 \pm 0.53 \times 10^{6}$ km $= 0.4998 \pm 0.0035$ AU.  Assuming $i = 67\degr$, $a \sim 0.54$ AU.

The distance to T CrB is also somewhat uncertain.  The {\it Gaia} Data Release 2 (DR2) parallax is reported as $1.21 \pm 0.05$ milliarcseconds \citep{gaiadr2}.  This translates to a nominal distance of $806^{+34}_{-30}$ pc \citep{bj18}. 
Note, however, that the {\it Gaia} DR2 parallax measurements assume that all sources are single stars and they do not account for any errors that are introduced by measuring the photocenter of a binary rather than the astrometric position of the star \citep{lindegren18}.  In the case of wide binaries, the change in the position of the photocenter can be comparable to the parallax.  As pointed out in \cite{schaefer18}, T CrB may not have a reliable {\it Gaia} DR2 parallax measurement due to this issue.  The  binary separation of $\sim0.54$ AU at a distance of 806 pc translates to an angular separation of $\sim0.67$ mas, or about half of the reported parallax.  It is therefore quite possible that the {\it Gaia} DR2 parallax was affected by the orbital motion of the stars, especially if the {\it Gaia} visits were unevenly distributed with respect to the orbital phase of the binary.  Nevertheless, the {\it Gaia} DR2 parallax measurement is the best distance estimate currently available for T CrB, and we will use it throughout this paper.

\begin{figure}
\includegraphics[trim=0 0 10 30,clip,width=3.3in]{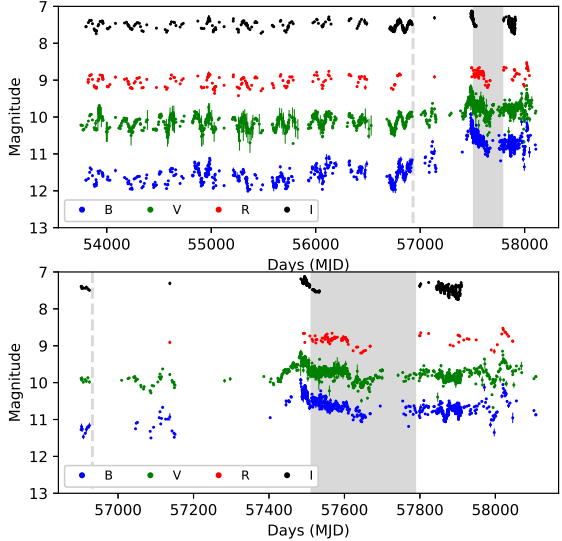}
\caption{The BVRI light curves for T CrB.  All data obtained from AAVSO.  Gray regions indicate the timespan of our VLA observations.  Vertical dashed gray line indicates the single 2014 VLA observation.  {\bf Top:} From 2006-01-01 to 2017-12-20.  {\bf Bottom:} From 2014-09-01 to 2017-12-20.}
\label{tcrb_aavso}
\end{figure}

During 2015 and 2016, T CrB appeared to be in a high-activity state \citep{munari16, tcrb_flickering16, zamanov_atel}.  Light curves obtained from the American Association of Variable Star Observers (AAVSO) are shown in Figure~\ref{tcrb_aavso} \citep{aavso18}.  Photometric optical monitoring during this time showed a significant increase in the mean brightness and revealed that the normal orbital modulation disappeared in the \emph{B}-band, while spectroscopic monitoring by \cite{munari16} showed new \ion{O}{4} and [\ion{Ne}{5}] emission lines.  The donor star spectral type also seemed to change, looking more similar to a M2III star on the side facing the white dwarf \citep{munari16}.  It has been suggested that this is evidence for a new ``super-active'' state that may precede a nova eruption in T CrB by approximately 8 years, from comparison to observations in 1938 before the 1946 nova eruption \citep{munari16}.  To explain the origin of this ``super-active'' state, \cite{munari16} hypothesize a large increase in emission from the ``hot source'' (presumably the accretion disk around the white dwarf) causing increased ionization of the donor wind to the point where it is nearly 100\% ionized.  X-ray and UV observations by \cite{lunaxray18} and \cite{luna19} provide direct evidence for increased ionizing radiation from the central source.

The peak of the ``super-active'' photometric brightening came in 2016 April.  Even after the peak, the source remained above the quiescent levels.  After the optical peak, \cite{lunaxray18} reported that the X-ray emission also declined and softened, with the hardness ratio softening by a factor of $\sim100$ due to the emergence of a new soft component.  They note that these changes in the X-rays could be attributed to a change in optical depth of the boundary layer after an increase in accretion rate.  The higher accretion rate also explains the increase in UV flux reported in \cite{lunaxray18} and the optical brightening.

However, \cite{tcrb_flickering16} suggest that this ``super-active'' state is merely a slightly higher manifestation of the known active phases of T CrB.  They base their analysis primarily on the equivalent width of the H$\alpha$ line, but also note that the \cite{munari16} photometric variability is consistent with previous ``big'' active phases.  It should be noted that \cite{tcrb_flickering16} do not address the lack of orbital modulation in the \emph{B}-band or the apparent change in spectral type of the donor star.

Because the radio emission of a symbiotic system is produced by thermal bremsstrahlung, radio observations are useful for obtaining information about the ionization of the surrounding material.  Here, we present our radio observations of T CrB from 2016 May to 2017 February.  The details of the observations are presented in Section~\ref{observations}.  Our results are presented in Section~\ref{results}.  We discuss our interpretations of these results in Section~\ref{discussion} and conclude in Section~\ref{conclusions}.

\section{Observations}
\label{observations}

We observed T CrB with the VLA between 2016 May and 2017 February at L, C, X, Ku, and Ka bands.  The L band observations were performed in 8-bit mode with 1 GHz of bandwidth.  The C, X, Ku, and Ka band observations were performed in 3-bit observing mode with 4 GHz of bandwidth at C and X bands, 6 GHz of bandwidth at Ku band, and 8 GHz of bandwidth at Ka band.  All bands were split into two sidebands (low and high) to maximize the spectral coverage of these observations.  All observations were performed under the program code VLA/16A-258.  These observations spanned three VLA configurations: CnB, B, and A.  The final observation was performed during ``move time'' when the VLA was transitioning between A and D configuration.  For each observation, we used 3C286 as the absolute flux calibrator.  Pointing solutions were obtained on 3C286 and J1602+3326 and applied to the Ku and Ka band observations.  Phase solutions were obtained with scans on J1609+2641 at L, C, and X bands for all configurations.  For Ku and Ka bands, phase solutions were obtained on J1602+3326 during CnB and B configurations, and J1619+2247 during A configuration.  

In order to obtain deep, high angular resolution images of the system during A configuration, for the observation on 2016 December 19 we omitted the L and C band observations in favor of longer X, Ku, and Ka band observations.  The aim was to look for any extended structure in the system or multiple components that would indicate an ejection event or possible jet structure.  The results of these observations are presented in Section~\ref{radio_size}.

Calibration for each observation was done with the VLA scripted calibration pipeline written\footnote{science.nrao.edu/facilities/vla/data-processing/pipeline/scripted-pipeline} for CASA \citep{casa}.  The calibrated data were then imported in AIPS \citep{aips} for additional flagging.  The flagged data were then imported into Difmap \citep{difmap} for imaging.  None of the observations had sufficient signal-to-noise for reliable self-calibration.  The observations are summarized in Table~\ref{radiotab}.  \cite{pb13} report that the VLA absolute flux density calibration is stable to within $1\%$ for 1 to 20 GHz, and within $3\%$ for 20 to 50 GHz.  Because T CrB is more than 10 degrees away from 3C286, we adopt a more pessimistic absolute flux density uncertainty of $3\%$ for 1 to 12 GHz and $5\%$ for above 12 GHz to account for variations in atmospheric conditions.  These flux uncertainties were added in quadrature to the image rms uncertainties provided by the AIPS task \verb|JMFIT| to create a final measurement uncertainty.  For nondetections, we calculate the upper limit by adding the flux density value at the source location to three times the uncertainty.  To be as conservative as possible, if the flux density at the source location is negative, we take it to be zero. 

The observations in 2016 August and September were obtained during the time when the VLA atmospheric delay model problem caused issues with source location and smearing along the direction of elevation\footnote{science.nrao.edu/facilities/vla/data-processing/vla-atmospheric-delay-problem}.  The effect caused by this issue was worse for higher frequencies and at lower elevations.  To ensure that our flux densities were not impacted by this issue, we double-checked our 2016 August and September observations by recalibrating them with the VLA CASA pipeline version 1.3.11, which automatically corrects for this issue.  The flux densities we measured using the recalibrated data were nearly identical to the original measurements, so we determined that this known issue had no effect on our measured flux densities.

While the 2016 August and September observations were not impacted by the atmospheric delay model problem, the 29.5 and 25.0 GHz observations were found to be decorrelated.  For our observations above 12 GHz, we also observed J1609+2641 as a ``check source''.  This bright source allowed us to determine if the phase solutions from the complex gain calibrator were properly applied to the observed sources.  Unfortunately, the check source required self-calibration to fully recover the total flux density.  We have adjusted the measured 29.5 and 35.0 GHz flux densities of T CrB on 2016 August 25 and 2016 September 22 by increasing them by the same percentage that the self calibration increased the flux density of J1609+2641.  We also increased the uncertainties of those measurements accordingly.

\startlongtable
\begin{deluxetable}{ccccccc}
\tablewidth{0 pt}
\tabletypesize{\footnotesize}
\setlength{\tabcolsep}{0.025in}
\tablecaption{ \label{radiotab}
VLA Observations}
\tablehead{UT Date & MJD & VLA & Freq. & $\Delta t$ & $S_{\nu}$ & $\sigma_{S_{\nu}}$ \\
&  & Config & (GHz) & (minutes) & (mJy) & (mJy) }
\startdata
2014-10-02 & 56932.02 & DnC & 10.0 & 95.7 & 0.040 & 0.004 \\
2016-05-04 & 57512.39 & CnB & 1.26 & 7.7 & \textless0.609 & 0.203 \\
 & & & 1.74 & 7.7 & \textless0.330 & 0.110 \\
 & & & 5.0 & 7.7 & \textless0.0891 & 0.017 \\
 & & & 7.0 & 7.7 & 0.080 & 0.013 \\
 & & & 9.0 & 7.2 & 0.131 & 0.013 \\
 & & & 11.0 & 7.2 & 0.148 & 0.015 \\
 & & & 13.5 & 6.8 & 0.192 & 0.015 \\
 & & & 16.5 & 6.8 & 0.215 & 0.016 \\
 & & & 29.5 & 6.5 & 0.396 & 0.033 \\
 & & & 35.0 & 6.5 & 0.441 & 0.037 \\
2016-06-04 & 57543.33 & B & 1.26 & 7.6 & \textless0.568 & 0.019 \\
 & & & 1.74 & 7.6 & \textless0.297 & 0.074 \\
 & & & 5.0 & 7.7 & 0.043 & 0.014 \\
 & & & 7.0 & 7.7 & 0.043 & 0.014 \\
 & & & 9.0 & 7.3 & 0.154 & 0.013 \\
 & & & 11.0 & 7.3 & 0.179 & 0.015 \\
 & & & 13.5 & 7.2 & 0.217 & 0.016 \\
 & & & 16.5 & 7.2 & 0.251 & 0.018 \\
 & & & 29.5 & 6.8 & 0.443 & 0.032 \\
 & & & 35.0 & 6.5 & 0.485 & 0.037 \\
2016-07-14 & 57583.18 & B & 1.26 & 7.6 & \textless0.426 & 0.142 \\
 & & & 1.74 & 7.6 & \textless0.173 & 0.057 \\
 & & & 5.0 & 7.7 & 0.075 & 0.014 \\
 & & & 7.0 & 7.7 & 0.096 & 0.010 \\
 & & & 9.0 & 7.3 & 0.145 & 0.013 \\
 & & & 11.0 & 7.3 & 0.196 & 0.014 \\
 & & & 13.5 & 6.6 & 0.227 & 0.016 \\
 & & & 16.5 & 6.6 & 0.287 & 0.020 \\
 & & & 29.5 & 6.2 & 0.445 & 0.034 \\
 & & & 35.0 & 6.2 & 0.513 & 0.038 \\
2016-08-25 & 57630.58 & B & 1.26 & 7.6 & \textless0.701 & 0.175 \\
 & & & 1.74 & 7.6 & \textless0.324 & 0.080 \\
 & & & 5.0 & 7.7 & \textless0.112 & 0.022 \\
 & & & 7.0 & 7.7 & 0.077 & 0.010 \\
 & & & 9.0 & 7.3 & 0.157 & 0.015 \\
 & & & 11.0 & 7.3 & 0.177 & 0.015 \\
 & & & 13.5 & 7.0 & 0.206 & 0.015 \\
 & & & 16.5 & 7.0 & 0.265 & 0.019 \\
 & & & 29.5 & 6.6 & 0.243 & 0.031 \\
 & & & 35.0 & 6.6 & 0.306 & 0.035 \\
2016-09-22 & 57653.05 & B$\rightarrow$A & 1.26 & 7.5 & \textless0.363 & 0.121 \\
 & & & 1.74 & 7.5 & \textless0.178 & 0.053 \\
 & & & 5.0 & 7.6 & 0.044 & 0.012 \\
 & & & 7.0 & 7.6 & 0.154 & 0.016 \\
 & & & 9.0 & 7.3 & 0.138 & 0.014 \\
 & & & 11.0 & 7.3 & 0.161 & 0.015 \\
 & & & 13.5 & 9.1 & 0.219 & 0.016 \\
 & & & 16.5 & 9.1 & 0.244 & 0.018 \\
 & & & 29.5 & 9.3 & 0.358 & 0.034 \\
 & & & 35.0 & 9.3 & 0.332 & 0.037 \\
2016-12-19 & 57741.76 & A & 9.0 & 7.9 & 0.178 & 0.013 \\
 & & & 11.0 & 7.9 & 0.244 & 0.016 \\
 & & & 13.5 & 11.1 & 0.232 & 0.014 \\
 & & & 16.5 & 11.1 & 0.279 & 0.018 \\
 & & & 29.5 & 11.3 & 0.481 & 0.033 \\
 & & & 35.0 & 11.3 & 0.484 & 0.035 \\
2017-02-03 & 57787.69 & A$\rightarrow$D & 1.26 & 7.5 & \textless1.254 & 0.418 \\
 & & & 1.74 & 7.5 & \textless0.693 & 0.231 \\
 & & & 5.0 & 7.7 & \textless0.121 & 0.023 \\
 & & & 7.0 & 7.7 & 0.082 & 0.026 \\
 & & & 9.0 & 7.3 & 0.120 & 0.022 \\
 & & & 11.0 & 7.3 & 0.209 & 0.024 \\
 & & & 13.5 & 9.2 & 0.234 & 0.018 \\
 & & & 16.5 & 9.2 & 0.298 & 0.022 \\
 & & & 29.5 & 8.8 & 0.540 & 0.041 \\
 & & & 35.0 & 8.8 & 0.603 & 0.049 \\
\enddata
\end{deluxetable}

\newpage
In addition to our new VLA observations, we had access to a single X-band observation of T CrB from 2014 October 02.  This observation was done as part of the Ph.D. thesis of J.H.S.~Weston.  The flux density for this observation was $0.040 \pm 0.004$ mJy at a central frequency of 10 GHz, as reported in \cite{linford_atel}.  

We also analyzed archival (pre-ELVA-upgrade) VLA observations.  Most of these observations were part of programs lead by R. Hjellming.  The project codes for these observations were AT59, TEST, AT102, AS378, and AS430.  All of the archival observations were calibrated using standard procedures in AIPS.  For the archival observations we use a more conservative estimate of the absolute flux calibration uncertainty: 5\% for 1--12 GHz, and 10\% for frequencies above 12 GHz.  All of the archival observations were nondetections.  Again, flux calibration and image rms uncertainties were added in quadrature to obtain the final measurement uncertainty.  We calculated the upper limits in the same way as described above for the recent observations.  The archival results are summarized in Table~\ref{archivaltab}.  While the 4.86, and 14.94 GHz archival observations produced upper limits above the detected radio emission at comparable frequencies shown in Table~\ref{radiotab}, the 8.44 GHz archival upper limits are very close to the new 9.0 GHz measurements.

\vspace{24pt}

\section{Results}
\label{results}


\subsection{Size of Radio Photosphere}
\label{radio_size}

\begin{figure}
\includegraphics[trim=0 140 140 138,clip,width=3.3in]{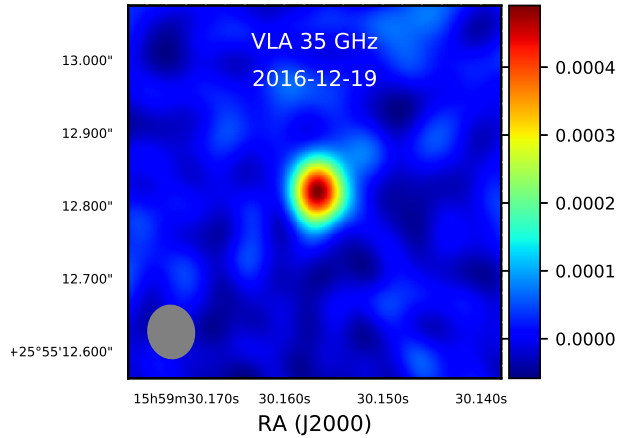}
\caption{The 35.0 GHz image of T CrB obtained on 2016-12-19.  We find no evidence for extended emission or jet structures.  The pixels are 2 mas square and natural weighting was used.  The gray ellipse in the lower left represents the restoring beam for this observation, which had an FWHM of 72.4 $\times$ 62.8 mas.  This was the highest resolution image obtained.}
\label{tcrb_highres}
\end{figure}

As mentioned in Section~\ref{observations}, we obtained a long-track observations of T CrB at higher frequencies (12--37 GHz) on 2016 December 19 during A configuration to look for extended structure.  The resulting images did not reveal any extended structure.  The source was unresolved at all frequencies.  The highest resolution image (see Figure~\ref{tcrb_highres}) had a restoring beam (effectively, the telescope resolution) of 72.4 $\times$ 62.8 mas.  Assuming the \cite{bj18} distance of 806 pc is accurate, this means the largest the radio photosphere could have been during that observation was $\sim50.8$ AU ($7.6\times10^{14}$ cm).

It is not surprising that the radio-emitting region was unresolved.  We can estimate the size of the radio photosphere from the brightness temperature ($T_{B}$):
\begin{equation}
T_{B} = \frac{2 \ln 2 \lambda^{2} S_{\nu} }{\pi k_{B} \theta_{d}^{2}}
\end{equation}
where $\lambda$ is the observing wavelength, $S_{\nu}$ is the observed radio flux density, $k_{B}$ is Boltzmann's constant, and $\theta_{d}$ is the angular diameter of the emitting region.  Solving for $\theta_{d}$, assuming that the radio-emitting material is all photoionized hydrogen at $T=10^{4}$K, and using our flux density measurements from 2016 December 19, gives an expected angular diameter in the range of 6.9--16.4 mas, well below our angular resolution.

Using the \cite{bj18} distance of 806 pc, our estimate of the radio photosphere angular diameter translates to a physical diameter in the range of $\sim$5.6--13.2 AU. This indicates that the radio photosphere should be well outside the binary separation of $\sim0.54$ AU (see Section~\ref{intro}).  In fact, using the \cite{fekel2000} $a \sin i = 0.4998 \pm 0.0035$ AU and our lower limit on the diameter of the radio photosphere of $\sim5.6$ AU, the radio photosphere would be larger than the binary separation for all $i\gtrsim 10\degr$.

Comparing the 10 GHz observation from 2014 October 02 with the 9 and 11 GHz observations on 2016 December 19, indicates that the radio photosphere has expanded significantly.  The photosphere diameter estimate from the 2014 observation was$\sim4.6$ AU, while the average of the diameter estimates for the 9 and 11 GHz observations in 2016 was $\sim12.9$ AU.  Therefore, assuming that the emission mechanism was the same for the two observations, the radio photosphere seems to have expanded by more than a factor of 2 in $\sim2$ years.

\subsection{Radio Light Curve}

\begin{figure}
\includegraphics[trim=0 0 10 30,clip,width=3.3in]{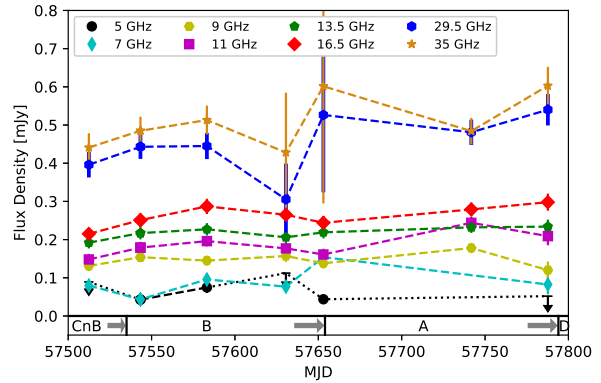}
\caption{The VLA radio light curve for T CrB from 2016 May 05 to 2017 Feb. 03.  Nondetections (3$\sigma$ upper limits) are indicated by downward-pointing triangles.  
The VLA configuration is indicated on the bottom; gray arrows indicate ``move time'' when the VLA is transitioning from one configuration to another.  The large error bars on the fourth and fifth 29.5 and 35.0 GHz observations are due to the decorrelation issues described in Section~\ref{observations}.}
\label{tcrb_lc}
\end{figure}

Our 2016 observations of T CrB revealed that the X-band (8--12 GHz) flux density has definitely increased since the previous detection in 2014.  We also see significant variability in the radio flux density, especially at Ka band (26.5--40 GHz).  As mentioned in the previous section, the variability at Ka band cannot be explained by the VLA atmospheric delay model problem.  The radio light-curve is shown in Figure~\ref{tcrb_lc}.

We wanted to rule out the possibility that any variability seen in the radio light curve was due to the orbit of the binary.  
We fit the radio light curve data with a periodic signal described by the Fourier series \citep[e.g., ][]{rucinski73, dey15}
\begin{equation}
S_{\nu}(t) = \sum_{i=1}^{2} [a_{i} \cos(2\pi t/P) + b_{i} \sin(2\pi t/P) ]
\end{equation}
where $S_{\nu}$ is the radio flux density at frequency $\nu$, $P$ is the known orbital period of the binary (227.6 days), and $t$ is time.  We then tested the goodness-of-fit of the modeled light curves to the data with reduced chi-squared statistics.  The resulting reduced chi-squared values for all frequencies were greater than 360.  We therefore conclude that the orbital motion has no discernible impact on the radio light curve.  This is the result we expected considering the estimated diameter of the radio photosphere in the previous section is at least an order of magnitude larger than the binary separation.  

\begin{deluxetable}{ccccccc}
\tablewidth{0 pt}
\tabletypesize{\footnotesize}
\setlength{\tabcolsep}{0.025in}
\tablecaption{ \label{archivaltab}
Archival VLA Observations}
\tablehead{UT Date & MJD & VLA & Freq. & $\Delta t$ & $S_{\nu}$ & $\sigma_{S_{\nu}}$ \\
&  & Config & (GHz) & (minutes) & (mJy) & (mJy) }
\startdata
1985-06-14 & 46230 & BC & 4.86 & 11.3 & \textless0.385 & 0.128 \\
 & & & 1.44 & 10.7 & \textless0.778 & 0.237 \\
1986-09-30 & 46703 & BC & 4.86 & 13.0 & \textless0.486 & 0.121 \\
 & & & 14.94 & 13.0 & \textless0.715 & 0.238 \\
 & & & 1.49 & 11.8 & \textless0.636 & 0.212 \\
1989-05-19 & 47665 & BC & 8.44 & 21.7 & \textless0.165  & 0.040 \\
1989-10-31 & 47830 & CD & 8.44 & 19.7 & \textless0.135  & 0.041 \\
1991-02-26 & 48313 & CD & 8.44 & 20.0 & \textless0.193  & 0.053 \\
\enddata
\end{deluxetable}

\subsection{Radio Spectral Index}

\begin{figure}
\includegraphics[trim=0 65 10 70,clip,width=3.3in]{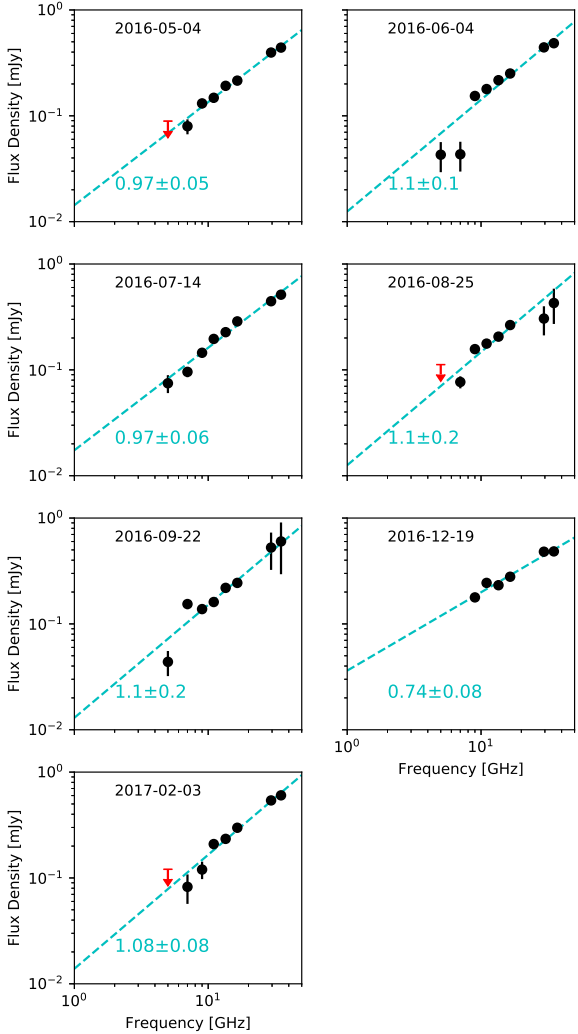}
\caption{Spectral energy distribution of T CrB from 2016 May 05 to 2017 Feb. 03.  Nondetections (3$\sigma$ upper limits) are indicated by downward-pointing red triangles.  The dashed cyan lines indicate the best (linear) fit for the spectral index.  The value for the spectral index is given in cyan.}
\label{tcrb_si}
\end{figure}

We measured the spectral index $\alpha$ (using $S_{\nu}\propto\nu^{\alpha}$) for each of our VLA observations.  We fit the flux densities to a power law using the nonlinear least squares \verb|curve_fit| function in the \emph{SciPy} package of python, weighted by the 1$\sigma$ uncertainties.  We ignored upper limit values for this fit, using only solid detections.  The spectral indices for the 2016 VLA observations are all positive (i.e., the flux density rises toward higher frequencies).  The values for $\alpha$ range from 0.74 to 1.08, with an average for the seven observations of $\alpha=0.98$. This is consistent with the spectral index expected for optically thick thermal bremsstrahlung emission.  The spectral energy distributions with the calculated spectral indices are shown in Figure~\ref{tcrb_si}.  Note that the calculated spectral indices do not account for the possibility of spectral curvature or a knee in the spectral energy distribution.  

\section{Discussion}
\label{discussion}


\subsection{Radio Variability}

Long-term radio variability in symbiotic binaries is not uncommon.  \cite{st90} reported that roughly 30\% of symbiotic systems were found to be variable on month to year timescales.  However, that study was conducted primarily at 15 GHz and below.  There is not a large amount of data about the radio behavior of symbiotic binaries at frequencies greater than 20 GHz.  The radio light curves presented in Figure~\ref{tcrb_lc} seem to show significant variability at multiple frequencies.

To quantify the variability in the light curves, we calculated the $V$ and $\eta$ parameters \citep[e.g.,][]{scheers11,stewart16}.  The $V$ parameter is the coefficient of variability and measures the amplitude of variability. $V$ is given by
\begin{equation}
V = \frac{1}{\langle S_{\nu} \rangle} \Bigg[\frac{N}{N-1} \Bigg(\langle S^{2}_{\nu}\rangle - \langle S_{\nu}\rangle^{2} \Bigg) \Bigg]^{1/2}
\end{equation}
where $S_{\nu}$ is the flux density, $N$ is the number of measurements in the sample, and $\langle \rangle$ indicates an arithmetic mean.  Larger $V$ values indicate larger variations in the light curve.
The $\eta$ parameter is a significance value (based on reduced $\chi^{2}$ statistics) and indicates how well a light curve is modeled by a constant value.  It is given by
\begin{equation}
\eta = \frac{N}{N-1} \Bigg( \langle wS_{\nu}^{2} \rangle - \frac{\langle wS_{\nu} \rangle^{2}}{\langle w \rangle} \Bigg)
\end{equation}
where $w$ is the weight which is the inverse-square of the flux density measurement uncertainty ($w=\sigma_{S_{\nu}}^{-2}$).  Larger $\eta$ values indicate more significant variability.  

\begin{figure}
\includegraphics[trim=0 10 0 10,clip,width=3.0in]{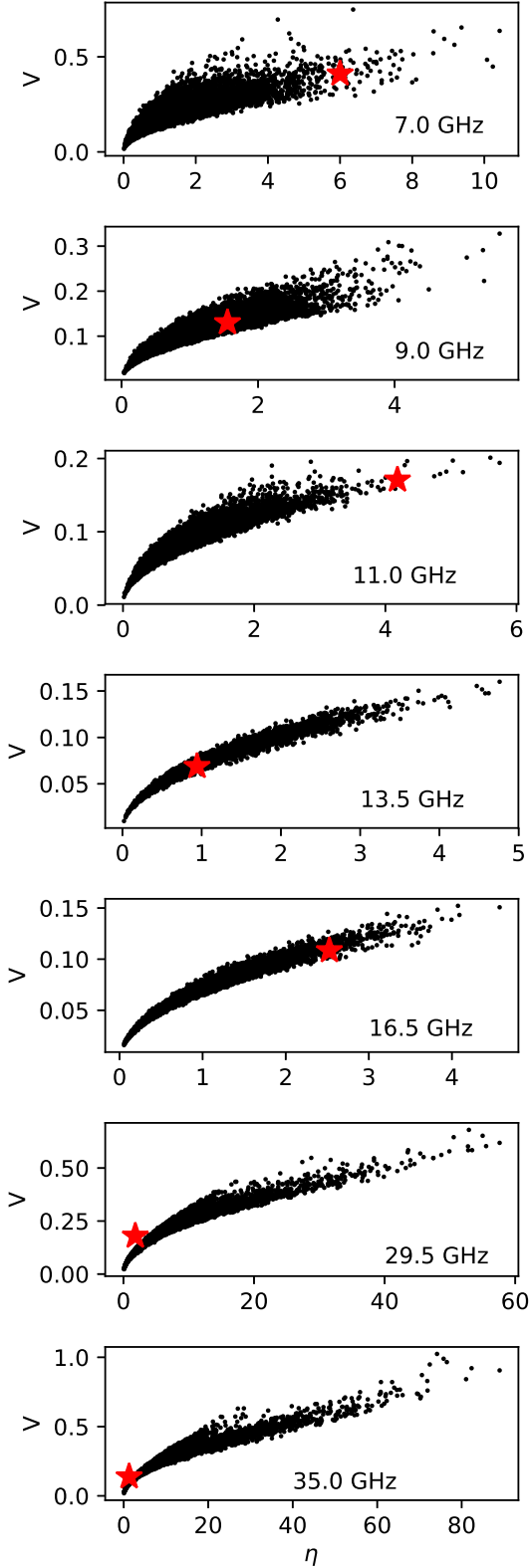}
\caption{The V and $\eta$ measurements for the radio light curves.  Values for the simulated light curves are shown as black points.  The values for the observed light curves are shown as red stars.}
\label{tcrb_eta_v}
\end{figure}

To test the significance of the measured $V$ and $\eta$ values of individual light curves, we created simulated radio light curves.  We began with the assumption that the radio emission is not variable, but rather the result of normal uncertainties in the measurements.  We therefore created synthetic light curves by randomly drawing numbers from a uniform distribution centered around the weighted mean of flux density of the observed light curve at each frequency.  To mimic the observed light curves as closely as possible, the standard deviation of each simulated flux density value (for given values of epoch and frequency) was the weighted mean of the light curve times the fractional uncertainty of the flux density from that observing epoch.  We created 10,000 synthetic light curves for each frequency from 7.0 to 35.0 GHz and calculated $V$ and $\eta$ for each one.  We did not simulated the 1.26, 1.74, or 5.0 GHz light curves because of the large number of upper limits at these frequencies.  The results of the simulations compared to the observed light curves are shown in Figure~\ref{tcrb_eta_v}.

The $V$ and $\eta$ values for the synthetic light curves do not follow Gaussian distributions.  Instead, they all have tails out to high values.  
For each frequency, we determined the number of times the $V$ and $\eta$ values for the synthetic light curves were below the values of the observed light curves.  These results are presented in Table~\ref{simtab}, and show how significant the variability was at each frequency.  

Our simulations produced mixed results.  The 7.0, 11.0, and 16.5 GHz observed light curves have $V$ and $\eta$ values greater than at least 95\% of the simulated light curves, indicating that there is significant variability at these frequencies.  The 9.0 and 13.5 GHz observed light curves are consistent with a flat light curve.  The 29.5 and 35.0 GHz light curves are unreliable tests of variability due to the large uncertainties in the flux densities for the two days which had decorrelated phases.  While it is possible that there was some short-term variability in the radio light curve, our results are inconclusive.

\begin{deluxetable}{ccc}
\setlength{\tabcolsep}{0.15in}
\tablecaption{ \label{simtab}
$V$ and $\eta$ \\ Simulations vs. Observed}
\tablehead{Freq  & $V$ &  $\eta$ \\
(GHz) & \%\textless Obs & \%\textless Obs }
\startdata
7.0 & 98.4 & 99.4  \\
9.0 & 78.3 & 80.6  \\
11.0 & 99.7 & 99.9 \\
13.5 & 50.3 & 53.7 \\
16.5 & 96.9 & 98.1 \\
29.5 & 57.0 & 20.4 \\
35.0 & 24.3 & 8.1 \\
\enddata
\end{deluxetable}

\subsection{Cause of Increased Radio Flux Density}

The increased flux density of T CrB during our observations compared to the 1985--2014 observations lends support to the theory that T CrB was in a state of increased activity during 2016 and 2017.  \cite{munari16} suggest that the system may have been undergoing increased accretion, leading to increased ionization in the surrounding material.  \cite{lunaxray18} provide additional evidence for an increased accretion rate because the observed UV and soft X-ray luminosity was much higher, while the hard X-ray luminosity declined, compared to earlier observations. These observations can be interpreted as due to an increase in the accretion rate through the inner disk (UV) and through the accretion disk boundary layer. The latter changed the boundary layer to become predominantly optically thick, reducing the hard X-ray luminosity and creating a previously unseen soft X-ray component. Independently of any such interpretations, the observed increase in the UV and soft X-ray luminosity would have caused more of the M giant wind to become ionized, dramatically strengthening the high excitation optical emission lines \citep{munari16}.

An increased accretion rate is a good explanation for the increased radio flux density because the radio thermal bremsstrahlung emission should originate from ionized material.  As more of the surrounding material is ionized, more material is available to produce radio emission.  
As shown in \cite{stb84}, \cite{ts84}, and \cite{st92}, the ionized region of the symbiotic system is expected to be optically thin at most GHz frequencies.  The shape of the ionized region is a function of the binary separation $a$, the mass-loss rate of the donor star $\dot{M}$, the velocity of the wind from the donor star $v$, and the ionizing luminosity of the white dwarf $L_{ph}$.  The location of the ionization front, as presented in \cite{ts84}, is given by
\begin{equation}
f(u,\theta) = \frac{4 \pi \mu^{2} m_{\rm H}^{2} v^{2}}{\alpha \dot{M}^{2}} a L_{ph} = X
\label{stb_eq}
\end{equation}
where $\mu$ is the reduced mass, and $\alpha$ is the recombination coefficient to all but the ground state of hydrogen.  In this notation, $u$ is the distance from the white dwarf to the ionization front in units of the binary separation (i.e., $u = r/a$), and $\theta$ is the angle measured from the binary plane (i.e., $\theta=0$ is the line pointing from the white dwarf to the companion star, $\theta=\pi / 2$ is a line normal to the binary plane centered on the white dwarf).  See \cite{stb84} and \cite{ts84} for details and illustrations.

Changes in the wind velocity $v$, mass-loss rate $\dot{M}$ and/or the ionizing luminosity of the white dwarf $L_{ph}$ change the shape and size of the ionized region around the binary\footnote{It should be noted that $v$, $\dot{M}$, and $L_{ph}$ are likely interconnected.  For example, increased $v$ and/or $\dot{M}$ should lead to increased accretion rate and therefore increased $L_{ph}$.  However, there are cases where $L_{ph}$ increases independent of $v$ or $\dot{M}$, such as accretion instability events (e.g., dwarf novae) or fusion on the surface of the white dwarf (e.g., novae).}.  \cite{ts84} show that for $X=1/3$, the wind is ionized to infinity along the line $\theta=\pi$ (i.e., along a line from the white dwarf pointing directly away from the companion star).  At $X=\pi /4$, the wind is ionized to infinity at all $\theta \geq \pi /2$.  Note that \cite{munari16} report that the wind is completely ionized along the line of sight, indicating $X \geq 1/3$.  The lack of significant variation in the radio light curve over one orbital period of the system (see Figure~\ref{tcrb_lc}) indicates that the ionized region extends beyond the donor star, thus $X > \pi/4$.

Neither \cite{munari16} nor \cite{lunaxray18} mention any evidence for an increase in the wind velocity in their spectral measurements of T CrB.  There is also no mention of any change in the mass-loss rate of the companion star.  Therefore, the only possible explanation for a change in the shape of the ionization front is a change in the ionizing luminosity of the white dwarf, $L_{ph}$.  Because we found that the size of the radio-emitting region was significantly larger during the ``super-active'' state, $L_{ph}$ must be larger as well.  This agrees well with the conclusions of \cite{munari16} and \cite{lunaxray18}.

\section{Conclusions}
\label{conclusions}
The symbiotic binary system T CrB had higher radio flux density state during 2016 and early 2017 compared to observations between 1985 and 2014.  
This state of higher radio emission corresponded to a period of increased activity in optical bands \citep[e.g.,][]{munari16} and changes in the boundary layer emission in X-rays \citep[e.g.,][]{lunaxray18}.  We searched for evidence of short-term ($\sim$month timescale) variability in the radio flux density, but our results were inconclusive.  The radio photosphere is located well outside the binary system.  The increased radio emission during the ``super-active'' state can be explained by an increase in the ionizing luminosity from the white dwarf and its associated accretion disk.  The increased accretion rate reported in \cite{lunaxray18} would be expected to drive higher temperatures and increase the ionizing luminosity, which makes it a plausible explanation for the increased radio emission.

T CrB is expected to undergo another nova eruption in the next decade.  Naively taking the difference between the 1866 and 1946 eruptions and assuming this is the duty cycle indicates the system should erupt in 2026.  B. Schaefer (private communication, 2019) argues that an eruption could occur as early as mid-2022, but is more likely to occur in mid-2023.  Both B. Schaefer and \cite{munari16} point out that T CrB was observed in a high-activity state during the summer of 1938, $\sim8$ years prior to the 1946 eruption, indicating that this increase in activity may be a precursor to a nova eruption.  
We strongly recommend an increased focus on T CrB at all wavelengths to monitor the system prior to, during, and after the next nova eruption.

\acknowledgements
We thank the anonymous referee for their insightful comments and criticisms, which led to an improved version of this manuscript.  We appreciate the useful conversations about T CrB, CVs, and accretion at the Stellar Remnants at the Junction meeting and the Conference on Shocks and Particle Acceleration in Novae and Supernovae.
We thank Dr. Bob Zavala and the USNO review board for their helpful comments on this manuscript.
We acknowledge support from NASA award NNX14AQ36G.
J.~L.~S.\ and J.~H.~S.~W.\ were funded in part by NSF award AST-1211778. L.~C.\ acknowledges support from a Cottrell Scholarship of the Research Corporation and NSF AST-1751874.
We thank the NRAO for the generous allocation of VLA time for our observations. The National Radio Astronomy Observatory is a facility of the National Science Foundation operated under cooperative agreement by Associated Universities, Inc.  
We acknowledge with thanks the variable star observations from the AAVSO International Database \citep{aavso18} contributed by observers worldwide and used in this research.  
This research made use of \emph{SciPy} \citep{scipy}, an open-source library of numerical routines for scientific computing available at http://www.scipy.org; \emph{NumPy} \citep{numpy}, the fundamental package for scientific computing with Python; \emph{APLpy} \citep{aplpy}, an open-source plotting package for Python hosted at http://aplpy.github.com; and \emph{Astropy} \citep{astropy1,astropy2}, a community-developed core Python package for Astronomy available at http://www.astropy.org.  The figures in this manuscript were made using the matplotlib 2D graphics package for Python \citep{matplotlib07}.\\

\facility{Karl G. Jansky VLA, AAVSO}

\software{AIPS \citep{aips}, difmap \citep{difmap}, CASA \citep{casa}, SciPy \citep{scipy}, NumPy \citep{numpy}, APLpy \citep{aplpy}, AstroPy \citep{astropy1,astropy2}, matplotlib \citep{matplotlib07}}


\begin{thebibliography}{}

\bibitem[Allen (1984)]{allen84} Allen, D.~A. 1984, PASAu, 5, 369
\bibitem[Astropy Collaboration (2013)]{astropy1} Astropy Collaboration 2013, \aap, 558, A33
\bibitem[Astropy Collaboration (2018)]{astropy2} Astropy Collaboration 2018, \aj, 156, 123
\bibitem[Bailer-Jones et al. (2018)]{bj18} Bailer-Jones, C.~A.~L., Rybizki, J., Fouesneau, M., Mantelet, G., \& Andrae, R. 2018, \aj, 156, 58
\bibitem[Bailey (1975)]{bailey75} Bailey, J. 1975, JBAA, 85, 217
\bibitem[Belczynski \& Miko\l ajewska (1998)]{bm98} Belczynski, K., \& Miko\l ajewska, J. 1998, \mnras, 296, 77
\bibitem[Dey et al. (2015)]{dey15} Dey, A., Deb, S., Subhash, K., et al. 2015, Journal of Undergraduate Research and Innovation, 1, 60, arXiv: 1502.06212
\bibitem[Fekel et al. (2000)]{fekel2000} Fekel, F.~C., Joyce, R.~R., Hinkle, K.~H., \& Skrutskie, M.~F. 2000, \aj, 119, 1375
\bibitem[Gaia Collaboration et al. (2018)]{gaiadr2} Gaia Collaboration, Brown, A.~G.~A., Vallenari, A., et al. 2018, \aap, 616, A1
\bibitem[Greisen (2003)]{aips} Greisen, E.~W. 2003, in Astrpphysics and Space Science Library, Vol. 285, Information Handling in Astronomy -- Historical Vistas, ed. A. Heck (Kluwer Academic Publishers), 109
\bibitem[I\l kiewicz et al. (2016)]{tcrb_flickering16} I\l kiewicz, K., Miko\l ajewska, J., Stoyanov, K., Manousakis, A., \& Miszalski, B. 2016, \mnras, 462, 2695
\bibitem[Kafka et al. (2018)]{aavso18} Kafka, S., Observations from the AAVSO International Database, https://www.aavso.org
\bibitem[Kenyon \& Garcia (1986)]{kg86} Kenyon, S.~J., \& Garcia, M.~R. 1986, \aj, 91, 125
\bibitem[Lindegren et al. (2018)]{lindegren18} Lindegren, L., Hern\'{a}ndez, J., Bombrun, A., et al. 2018, \aap, 616, A2
\bibitem[Linford et al. (2016)]{linford_atel} Linford, J., Weston, J., Chomiuk, L., et al. 2016, The Astronomer's Telegram, 9153, 1
\bibitem[Luna et al. (2018)]{lunaxray18} Luna, G.~J.~M., Mukai, K., Sokoloski, J.~L., et al. 2018, \aap, 619, 61
\bibitem[Luna et al. (2019)]{luna19} Luna, G.~J.~M., Neslon, T., Mukai, K., \& Sokoloski, J.~L. 2019, \apj, 880, 94
\bibitem[Luri et al. (2018)]{luri18} Luri, X., Broan, A.~G.~A., Sarro, L.~M., et al. 2018, \aap, 616, A9
\bibitem[Hunter (2007)]{matplotlib07} Hunter, J.~D. 2007, Computing in Science \& Engineering, 9, 90
\bibitem[McMullen et al. (2007)]{casa} McMullin, J.~P., Waters, B., Schiebel, D., Young, W., \& Golap, K. 2007, Astronomical Dara Analysis Software and Systems XVI, 376, 127
\bibitem[Munari et al. (2016)]{munari16} Munari, U., Dellaporta, S., \& Cherini, G. 2016, \na, 47, 7
\bibitem[M\"{u}rset \& Schmid (1999)]{ms99} M\"{u}rset, U., Wolff, B., \& Jordan, S. 1997, \aap, 319, 201
\bibitem[Oliphant (2007)]{scipy} Oliphant, T.~E. 2007, CSE, 9, 10
\bibitem[Perley \& Butler (2013)]{pb13} Perley, R.~A., \& Butler, B.~J. 2013, \apjs, 204, 19
\bibitem[Robitaillle \& Bressert (2012)]{aplpy} Robitaille, T., \& Bressert, E. 2012, APLpy: Astronomical Plotting Library inPython, Astrophysics Source Code Library, ascl:1208.017
\bibitem[Ruci{\'n}ski (1973)]{rucinski73} Ruci{\'n}ski, S.~M. 1973, \actaa, 23, 79
\bibitem[Schaefer (2010)]{schaefer10} Schaefer, B. 2010, \apjs, 187, 275
\bibitem[Schaefer (2018)]{schaefer18} Schaefer, B. 2018, \mnras, 481, 3033
\bibitem[Scheers (2011)]{scheers11} Scheers, L.~H.~A., 2011, PhD thesis, University of Amsterdam
\bibitem[Seaquist, Taylor, \& Button (1984)]{stb84} Seaquist, E.~R., Taylor, A.~R., \& Button, S. 1984, \apj, 284, 202
\bibitem[Seaquist \& Taylor (1990)]{st90} Seaquist, E.~R., \& Taylor, A.~R. 1990, \apj, 349, 313
\bibitem[Seaquist \& Taylor (1992)]{st92} Seaquist, E.~R., \& Taylor, A.~R. 1992, \apj, 387, 624
\bibitem[Sheperd (1997)]{difmap} Sheperd, M.~C. 1997, in ASP Conf. Ser. 125, Astronomical Data Analysis Software and Systems VI, ed. G. Hunt \& H.~E. Payne (San Francisco, CA: ASP), 77
\bibitem[Stanishev et al. (2004)]{stan04} Stanisehv, V., Zamanov, R., Tomov, N., \& Marziani, P. 2004, \aap, 415, 609
\bibitem[Stewart et al. (2016)]{stewart16} Stewart, A.~J., Fender, R.~P., Broderick, J.~W., et al. 2016, \mnras, 456, 2321
\bibitem[Taylor \& Seaquist (1984)]{ts84} Taylor, A.~R., \& Seaquist, E.~R. 1984, \apj, 286, 263
\bibitem[van der Walt et al. (2011)]{numpy} van der Walt, S., Colbert, S.~C., \& Varoquaux, G. 2011, CSE, 13, 22
\bibitem[Warner (1995)]{warner95} Warner, B. 1995, Cataclysmic Variable Stars (Cambridge Univ. Press)
\bibitem[Yudin \& Munari (1993)]{ym93} Yudin, B., \& Munari, U. 1993, \aap, 270, 165
\bibitem[Zamanov et al. (2016)]{zamanov_atel} Zamanov, R., Semkov, E., Stoyanov, K., \& Tomov, T. 2016, The Astronomer's Telegram, 8675, 1




\end{thebibliography}
\end{document}